**Analysis of a kink pair model applied to a Peierls mechanism in basal and prism plane slips in sapphire (α-Al$_2$O$_3$) deformed between 200°C and 1800°C**

M. Castillo Rodríguez[1,], J. Castaing[2], A. Muñoz[1], P. Veyssière[3] and A. Domínguez Rodríguez[1, a, *]

**Abstract**

A model based on dislocation glide controlled by the nucleation and propagation of kink pairs in a high Peierls stress crystal is revisited and modified to account for changes in dislocation densities and segment lengths with temperature and stress. It is applied to the critical resolved shear stress (CRSS) for basal and prism plane slip in sapphire (α-Al$_2$O$_3$). According to agreed-upon knowledge on dislocations in sapphire, basal slip and prism plane slip are modelled with undissociated and dissociated dislocations, respectively. In the latter case, partial dislocations move independently. Amongst a number of sets of fitting parameters, good fits between experimental and modelled CRSS's are obtained in the *long segment* limit over the whole range of temperatures by making use of physically sound parameters, including a stress dependence of the dislocation density.

[1] Departamento de Física de la Materia Condensada. Universidad de Sevilla. P. O. 1045. 41081, Sevilla (Spain).

[2] C2RMF, CNRS UMR 171, Palais du Louvre, 14 quai François Mitterrand 75001 Paris (France)

[3] LEM, CNRS UMR 104, ONERA, BP 72, 92322 CHATILLON Cedex (France)

[a] Corresponding author. Email address: adorod@us.es

[*] Fellow, American Ceramic Society.

1. **INTRODUCTION**

The plasticity of sapphire (α-Al$_2$O$_3$ single crystal) has been extensively studied over the past four decades. Basal $(0001)1/3<\bar{2}110>$ slip is the primary deformation system for temperatures above 700 °C[1-7] while $\{1\bar{2}10\}<\bar{1}010>$ prism plane slip is the easiest slip system below 700 °C[3-4]. Consistent with the very long $<\bar{1}010>$ Burgers vector for prism plane slip in sapphire, dislocations are always observed under a dissociated form into three collinear partials with Burgers vector $1/3<\bar{1}010>$.[8] Figure 1 shows the critical resolved shear stress (CRSS) measured by various authors between 200 and 1800 °C for the two slip systems.[1-7] Castaing et al.[2-6] have fitted the temperature dependence of the CRSS, σ, on test temperature, T, to an empirical law lnσ = A - BT, where A and B are constants (Fig. 1). The quality of the fit is, however, poor and the physical basis of the law uncertain.

The Peierls mechanism is often invoked to explain the plastic properties of sapphire deformed in basal and prism plane slip as reviewed in Ref. 8. Starting from the work of Hirth and Lothe[9] for dislocation glide by nucleation and motion of kink pairs, Mitchell et al.[10-12] have developed a model that accounts for the temperature dependence of the CRSS in materials with a high Peierls stress. Applied to sapphire and spinel, the model provides apparently good fits of experimental data. As for sapphire, however, a number of difficulties remain with the choice made of the equation and with parameters, some of which are hard to justify. Firstly, one may wonder why Mitchell et al.[10], henceforth MPH, restricted their analysis to the length dependent *short-segment* limit of the kink pair model; as discussed later, their assumption on the values of the segment length is questionable. Secondly, dislocations are assumed to be dissociated into $1/3<\bar{1}010>$ partials regardless of the slip system activated[10, 12]. This is in agreement with observations for prism plane slip[3, 8] not so, however, for basal slip where according to transmission electron microscopy (TEM)[8, 13] dislocations are not dissociated in the glide plane, an observation supported by stacking fault energy calculations[8, 14] and molecular dynamic simulation.[15] Finally, making use of the kink pair model, MPH[10] have found somewhat low activation energies (1.5 and 2 eV), together with unrealistically low dislocation densities, well below 1 m$^{-2}$.

By making a more adequate use of the kink pair model for basal and prism plane slip in sapphire, including physically well-founded parameters, the present work shows that the experimental dependence of the CRSS on temperature is consistent with a Peierls mechanism controlled dislocation glide through a wide temperature range. A brief discussion of dislocation properties in relation with crystal periodicity is given in the appendix to justify the various kink heights used.

## 2. KINK MECHANISM MODEL FOR DEFORMATION

We start from the model updated by Mitchell *et al.*[11] In materials with high Peierls stresses, the rectangle-shaped kinks are abrupt and their motion is controlled by a secondary Peierls barrier. In the *long-segment* limit, the two kinks forming a pair move apart until they annihilate with kinks engendered by vicinal sources along the line (the so-called kink collision regime). In this case, Mitchell *et al.*[11] obtain for the strain rate:

$$\dot{\varepsilon} = \rho b a \upsilon \frac{\sigma' b_1 h^2}{\alpha^{1/2} kT} \exp\left(-\frac{Q'_D + F'_k}{kT}\right) \exp\left(\frac{(\sigma' b_1^3 h^3 \mu)^{1/2}}{(8\pi)^{1/2} kT}\right) \quad (1)$$

where $\rho$ is the mobile dislocation density, $b$ is the Burgers vector of the perfect dislocation, $b_1$ the partial Burgers vector, $h$ the kink height (i.e. the periodicity of the primary Peierls potential), $a$ the periodicity of the secondary Peierls potential, $\upsilon$ the attempt frequency, $\mu$ the shear modulus, $Q'_D$ the height of the secondary Peierls energy barrier, $F'_k$ the free energy of a single kink on a partial dislocation, $k$ the Boltzman constant, $T$ the absolute temperature and $\alpha$ a factor given by

$$\alpha = \frac{1}{2}\left\{1 + \left[1 + \left(\frac{2\sigma'\mu b_1^3 h^3}{\pi k^2 T^2}\right)^{1/2}\right]^{1/2}\right\} \qquad (2)$$

which is unity at high temperatures and more than unity for high stresses and low temperature. $\sigma'$ is the effective stress on the portion of the leading partial bordered by the kink pair

$$\sigma' = \sigma - \frac{\pi \gamma^2 h}{\mu b_1^3} \qquad (3)$$

where $\sigma$ is the resolved applied stress and $\gamma$ the stacking-fault energy (SFE).[11]

In the *short-segment* limit, the dislocation portion on which the kink pair nucleates is short enough to allow kink pair expansion over the full segment length $L$ before it annihilates with a colliding kink. In this case, the strain rate is given by

$$\dot{\varepsilon} = \rho b L \upsilon \frac{\sigma' b_1 h^2}{\alpha k T} \exp\left(-\frac{Q'_D + 2F'_k}{kT}\right) \exp\left(\frac{(\sigma' b_1^3 h^3 \mu)^{1/2}}{(2\pi)^{1/2} kT}\right) \qquad (4)$$

The segment length $L$ refers to mobile dislocations, hence the choice made in MPH[10] of $L = \rho^{-1/2}$, a relation commonly utilized for 3 dimensional dislocation networks within the well-known frame of the forest dislocation hardening. One may wonder though whether this expression is valid at the beginning of plastic deformation by basal slip in sapphire in which case, the dislocations glide on parallel planes and therefore cannot form forest-like obstacles and ensuing junctions. MPH[10] have ascribed $\rho$ a given value regardless of experimental conditions. In addition to clearly not being in agreement with observations within the 200-1800°C temperature range[1, 3, 7, 12], the assumptions made and the fitting conditions used by MPH[10] actually led them to utilize dislocation densities several orders of magnitude too small. More precisely, from the data in Table 1 of MPH[10], one finds dislocation densities of 0.4 m$^{-2}$ and 6 10$^{-6}$ m$^{-2}$ for basal and prism plane slip, respectively.

The situation greatly improves if one considers instead stress-dependent dislocation densities to account for high and low temperature regimes. To refine this aspect, we have postulated the dislocation density to be proportional to the square of the resolved shear stress:

$$\rho = \left(\frac{\sigma}{C\mu b_1}\right)^2 \quad (5)$$

which is merely the reciprocal of the well-known relation $\sigma = C \mu b_1 \rho^{1/2}$, where C is a constant. In other words, the dislocation density is directly related to the stress whereas, at variance from Ref. 10, the segment length $L$ is regarded as an independent parameter since there is no clear relationship between $\rho$ and $L$ at least at the beginning of deformation; for simplicity, we take constant $L$ values over the whole temperature range adjusted for each fit.

Equations (1) and (4) correspond to the case of glide controlled by a partial dislocation in the *long-* and *short-segment* regimes, respectively. In the limit defined by $b_1 = b$, where $b$ is the Burgers vector of the perfect dislocation, and $\sigma' = \sigma$, expressions (1) and (4) correspond to the case of kink pair nucleation on an undissociated dislocation, given that in this case $Q_D$ and $F_k$ have the same meaning as $Q'_D$ and $F'_k$.

Differences between equations (1) and (4) appear in both the pre-exponential and exponential terms. The ratio of the pre-exponential terms of the *short* to the *long-segment* limit is about $L/a$, that may take large values between 100 and 1000. On the other hand, the activation energy and the stress term are smaller in the *long-segment* limit than in the *short-segment* limit. A possible transition between the *short-segment* limit (low kink density regime) and the *long-segment* limit (kink collision regime) should be reflected by changes in activation energy and in the stress dependence of the deformation rate. This is, however, not confirmed experimentally since, as shown in section 3, the plastic behavior of sapphire can be fitted *graphically* over the whole temperature and stress range (Fig. 1) with no need to switch from Eq. (1) to Eq. (4).

## 3. COMPARISON BETWEEN EXPERIMENTAL DATA AND PREDICTIONS FROM THE KINK PAIR MODEL

### 3.1 Fitting procedure

In order to compare the data points (Fig. 1) to the CRSS values predicted by the kink pair model, a numerical fit by the analytical relations $\dot{\varepsilon}(\sigma, T; \mathbf{p})$ has been performed considering either the *long segment* limit (Eq. 1) or the *short segment* limit (Eq. 4). We define $\mathbf{p} = \{p_i\}$ as the set of parameters that can be adjusted within a fit. In the *long segment* limit $\mathbf{p} \equiv \{\mu, C, \rho, Q_D+F_k\}$ and in the *short segment* limit $\mathbf{p} \equiv \{\mu, C, L, \rho, Q_D+2F_k\}$. The elastic modulus $\mu$, which is a physical constant, has been either ascribed its experimental value at 20°C of 156 GPa or else adjusted, but kept reasonably close to its value. For the sake of simplicity, fixed values of $h$ and $a$ are used; $h = 0.275$ nm (oxygen – oxygen distance in the basal plane) for basal slip, $h = 0.65$ nm for prism plane slip and $a = 0.275$ nm for both slip systems. These values are compatible with the structure of sapphire (see appendix). The other parameters $b$, $b_1$ and $\gamma$ were taken from the literature,[3, 6, 8] while the parameter $\nu$ was ascribed as a fixed value. A least squares fitting routine has been used to determine the values for **p**. It is based on a Levenberg Marquardt algorithm provided in Mathematica™ by the subroutine Findfit. The sum of the squares of the deviations after minimization and the plots themselves are used to select the best fits amongst many attempts.

## 3.2   Basal plane slip

The values adopted are $\dot{\varepsilon} = 2 \cdot 10^{-5}$ s$^{-1}$, $\nu = 10^{13}$ s$^{-1}$, $b = 0.475$ nm (perfect dislocation) and $h = a = 0.275$ nm (see appendix). In Table I, we show the parameters of selected satisfactory fits of the experimental CRSS. In the *long-segment* limit, the commonly accepted value for $\mu = 156$ GPa yields large $\rho$ values (fit I, Table I) especially at low temperatures. If on the other hand, one allows some flexibility on $\mu$ (fit II), then one obtains dislocation densities realistically closer to experimental values.[16] Fit II (Table I and Fig. 2) yields an activation energy of $Q_D + F_k = 3.4$ eV and a value for $\mu$ not prohibitively inferior to its true value. The above flexibility on $\mu$ can be justified by the fact that the anisotropy and the temperature dependence of the elastic shear modulus are ignored. Notice that the fit of the experimental data based on Eq. (1) is excellent except above 1900 K (Fig. 2).

We have checked that fitted with $L = \rho^{-1/2}$, the *short-segment* limit implies physically unsound parameters since the values of both the activation energy and $\mu$ are too small while the dislocation density ($\rho \approx 4$ to $10^5$ m$^{-2}$) is in too severe a disagreement with experiments.[16] Although they had concluded to a fair agreement, MPH[10] were actually faced with a similar kind of discrepancy. As mentioned earlier (section II), we have assumed fixed $L$ values within the *short-segment* limit. Two results are shown in Table I with $L = 8.9$ µm (fit III) and $L = 1.2$ µm (fit IV), the latter corresponding to the plot in Fig. 3; the fit quality is, however, not as good as for that shown in Fig. 2. Obviously, it is not sound to use a unique value for $L$ over the whole range of experimental conditions (Fig. 1) but the assumption $L = \rho^{-1/2}$ (Ref. 10) is not adequate either. In all these fits, the values found for the dislocation density $\rho$ provided by Eq. (5) and for the activation energy are all reasonable (Table I). Hence, the fits obtained via the *short-segment* limit are nearly as good as those obtained with the *long-segment* limit (Figs. 2 and 3).

### 3.3   Prism plane slip

As mentioned earlier, prism plane slip proceeds by glide of widely dissociated dislocations.[3, 7, 8, 12] Following MPH[10], we therefore assume that deformation is controlled by the nucleation of a kink pair on the leading partial. The values adopted for the different parameters are $\dot{\varepsilon} = 2 \cdot 10^{-5}$ s$^{-1}$, $b = 0.822$ nm (perfect dislocation), $b_1 = a = 0.275$ nm (Burgers vector of the partial dislocation), $h = 0.65$ nm (see appendix) and $\gamma = 0.2$ J/m² (distances between partial dislocations about 30 nm). Similarly to what we have found for basal slip, reasonable fits cannot be obtained in the whole range of temperatures and stresses (Fig. 1) within both the *long-* and *short-segment* limits when one makes use of a constant density of mobile dislocations. Certain parameters associated with these fits do not have a reasonable physical justification; activation energies $Q'_D + F'_k$ (Eq. 1) or $Q'_D + 2F'_k$ (Eq. 4) are small (2.5 and 4.4 eV, respectively), so are the density $\rho$ of mobile dislocation values ($10^8$ and $10^9$ m$^{-2}$, respectively). The results with a constant $\rho$ correspond to those of MPH[10] obtained in the *short-segment* limit.

Similarly to what we have found for basal slip, the fits with a stress-dependent $\rho$ of the type given by Eq. (5) are satisfactory (fits I and II, Table II). The values for $\rho$ (Table II) seem excellent; it should be kept in mind though that there is no systematic experimental measurements of dislocation densities for prism slip to compare with. The fits yield reasonable values for the activation energies (Table II). So far, fit I and fit II (Table II), shown in Figs. 2 and 3 respectively, are the most satisfactory beside a significant discrepancy around 800-1000K between the experimental points and the predictions of Eq. (4) (fit II, Fig. 3). Therefore, as for basal slip, the *long segment* limit provides the best description of the temperature dependence of the CRSS in prism plane slip (Fig. 2), except at the extremes of the temperature range (below 500 K and above 1900 K).

## 4. DISCUSSION

As indicated by the above plots and tables, dislocation glide controlled by the nucleation and motion of kink pairs in a Peierls potential reasonably accounts for the plastic yielding of sapphire deformed either on basal or prism planes, over a large temperature range (200°C to 1800°C). Before discussing the fitting to the CRSS data of sapphire to the kink pair model, it is important to remember that the model is based on elasticity[9] applied to dislocations in the form of long as well as of very short (kinks) segments. This assumption, which is valid as long as the distances between interacting dislocation segments are larger than a few Burgers vector, may actually fail for exceedingly small *critical* separations between kinks, that is, at high stresses and low temperatures (the *critical* kink pair separation is the distance between the kinks above which the driving force encourages further kink separation[9]). On the other hand, whereas on account of the Peierls potential the kinks are supposed to assume a rectangular shape, thermal activation helps round the kink shape thus introducing a restriction in applying the kink pair model at high temperature. It is noticed that fit quality does not seem to suffer from these limitations except above 1600ºC (Fig. 2).

At variance with the work of MPH[10], the present analysis has been conducted under both the kink-collision *long-segment* limit and the length-dependent *short-segment* limit, with the further assumption that the segment length $L$ is either fixed or else defined by $L = \rho^{-1/2}$, itself related to the applied stress $\sigma$ via Eq. (5). The use of Eq. (5) may look natural since this is the relationship commonly observed between $\rho$ and the flow stress, for instance in the work-hardening stage of sapphire deformed by basal slip;[16] the values for $C$ (Eq. 5), given in Tables I and II, are close to those deduced from dislocation counting in transmission electron microscopy (0.2 – 0.5).[16] However, it is not clear whether or not Eq. (5) can be applied at the lower yield stress after the yield peak as is done in the present work. The model which Alexander and Haasen had designed to explain the yield peak behaviour in silicon, has been recently reformulated and checked by mesoscale simulation.[17] In brief, the Alexander-Haasen model takes into account the mobile dislocations with long range elastic interactions, at the origin of an internal stress proportional to $\rho^{1/2}$. Contrary to what is done in the Alexander-Haasen model, it is, however, not necessary to make use of an empirical equation in order to account for dislocation multiplication rate (see appendix in Ref. 17). A reasonably simple calculation shows that the stress at the lower yield point is indeed proportional to $\rho^{1/2}$ and independent of the initial dislocation density. This has also been checked by mesoscale modeling and is in agreement with the experimental results in silicon.[17] Making use of Eq. (5) for sapphire is therefore a reasonable way to reflect the expected variations of dislocation densities for deformations under shear stresses between 2500 MPa at 200°C and 7 MPa at 1600°C.[3-5]

As mentioned earlier, validating the kink-pair model relies in addition on how meaningful the fitting parameters may be in terms of certain physical characteristics that have been either observed or guessed. As regards $\nu$, $h$ and $a$, one is left with reasonable guesses. The frequency factor $\nu$ is the attempt frequency for the diffusion of a kink along the dislocation line; it is assumed to be of the order of the Debye frequency.[9] The Debye temperature of sapphire is about 1000 K giving a Debye frequency of $2 \cdot 10^{13}$ s$^{-1}$ consistent with our choice for $\nu = 10^{13}$ s$^{-1}$. As for the lattice periodicity $h$ for dislocation glide, it can be as large as the unit vectors in the slip planes, i.e. as much as 1.3 nm. In the appendix, smaller values are suggested that depend on the dislocation under consideration; they are close to $h = 0.275$ nm, the oxygen-oxygen distance in the basal plane. The fits for prism plane slip are made with $h = 0.65$ nm (Table II). The choice of $h$ is indeed crucial for it enters the argument of the exponential (Eqs. (1) and (4)). On the other hand, for the crystal potential oscillations that govern kink motion, it is difficult to suggest $a$ values based on a simple examination of the structure (see appendix), but this is not crucial because $a$ appears only in the pre-exponential term of Eq. (1). Worthy of mention is also the fact that while it is safe to follow MPH[10] in considering dissociated dislocations for prism plane slip (Table II), one should rather make use of *perfect* dislocations to account for basal slip (Table I).

The density $\rho$ of mobile dislocation is found between $1.9 \cdot 10^{11}$ m$^{-2}$ (Table II) and $3.8 \cdot 10^{16}$ m$^{-2}$ (Table I); these values correspond respectively to high temperature ($\sigma = 10 - 70$ MPa) and low temperature ($\sigma = 700 - 800$ MPa) limits of the investigated range. Not surprisingly they are at the border of what is acceptable according to the commonly accepted situation for plastic deformation, especially for basal slip at low temperature (Table I). It should be noted in Eqs. (1) and (4) that $\rho$ is related to the values of $\nu$, $h$ and $a$ that are "reasonably guessed"; accordingly, the uncertainty on $\rho$ values should be at most a factor of 10. The comparison with experimental data is made difficult by their scarcity and by the difficulty of counting dislocations due to their heterogeneous distribution, their annihilation after the deformation test has stopped, *etc.* In the conditions of the experiments (Fig. 1), the values for $\rho$ are reasonable except, possibly, at low temperature for basal slip.

As regards the two contributions to the activation energies, $Q_D$ and $F_k$, MPH[10] could not reach reasonably large values for $Q_D$ in both basal and prism plane slip. It is therefore interesting to examine the outputs of our fits in this respect. In the *long-segment* limit, the fitted activation energy ($Q_D + F_k$) for slip in the basal plane (3.5 eV; Table I) is similar to that in the prism plane (Table II). The values for the activation energy ($Q_D + 2F_k$) are larger within the *short-segment* approximation (Tables I and II). The values of $2F_k$ may be estimated directly from dislocation theory[9]; it is of the order of ($\mu\, b^2\, h\, /\, 4\pi$). Taking $\mu\, b_1{}^2 = 11$ nJ/m for a partial dislocation in the prism plane and $\mu\, b^2 = 34$ nJ/m for a perfect dislocation in the basal plane[8] (consistent with the molecular dynamic simulation of 48 nJ/m of the dislocation line energy[15]), one finds that for kinks of length $h = 0.65$ nm, $2F_k$ is 3.5 eV (prism plane) and for $h = 0.275$ nm, $2F_k$ is 4.6 eV (basal plane). Using the activation energy values from the fits in the *long-segment* limit (Tables I and II) yields $Q_D = 1.7$ eV (prism plane) and 1.2 eV (basal plane) that seem reasonable. The activation energies for basal and prism plane slip are therefore consistent with the model in the kink-collision *long-segment* limit. For basal slip in the length-dependent *short-segment* limit, the value of $Q_D$ would be close to zero which is not acceptable. Unfortunately, experimental activation energies are essentially nonexistent in sapphire[24] which makes it useless to further discuss this parameter.[25]

In the length dependent *short-segment* limit, the quality of the fits depends weakly on the value of $L$ and we arrive at $L \approx 1$ µm for basal and prism plane slips (Tables I and II). This value agrees with the typical lengths of dislocations observed in TEM after basal deformation at 1000°C (Ref. 18) and 800°C (Ref. 3) and after prism deformation at 400°C (Ref. 19). Small $L$ values are expected at high temperature because of an increased probability of cross-slip and climb events. Cadoz et al.[20] have reported that dislocations with lengths larger than 5 to 10 µm are currently observed in prism plane after deformation at 1400-1450°C, similarly to observations made in the basal plane.[1] Since the samples were not quenched, this inconsistency might, however, stem from high temperature recovery after the tests. There is nevertheless a certain lack of evidence in support of the length dependent *short-segment* limit to describe the temperature dependence of the CRSS in sapphire.

The kink-pair Peierls model has been invoked for dislocation glide in silicon with which sapphire exhibits strong similarities at low temperature[4, 5, 21]. Silicon being *the* model covalent crystal, dislocations have been the focus of many experimental and theoretical studies in this crystal, as reviewed by Kubin and Devincre[22]. A major property is that whereas dislocation velocity depends on segment lengths (Eq. 4), the transition to a length-independent regime is not associated with a change in the activation energy, thus ruling out the existence of a *long-segment* limit in terms of Eq. (1). Moreover, dislocation velocity in silicon can only be quantitatively accounted for by the length-dependent *short-segment* regime.[22] This suggests in turn an extrinsic transition to a length-independent regime corresponding to kink motion itself governed by the presence of defects on the dislocation lines.[22] Such a situation can be a priori envisaged for sapphire. However, not only does the *long-segment* limit fits best the CRSS values of sapphire doped with various Cr concentrations (rubies) but also this is achieved with fitting parameters that are essentially independent of the amount of Cr.[18] This latter property rules out a direct analogy with the case of silicon thus providing an additional argument in favor of the kink-collision *long-segment* limit to account for the deformation of sapphire.

## 5. CONCLUSIONS

Several sets of parameters can provide equally good fits between experimental data and predictions from a model. However, a necessary condition to end with an acceptable fit is to make sure that the various parameters entered in the fit are physically sound. We have shown that over the wide range of stresses considered (from less than 10 to more than 2000 MPa), this implies in practice that the dependence of $\rho$ on stress be accounted for. The present work supports the *long-segment* limit for basal and prism plane slip. The parameters included in the analysis of MPH[10], restricted to the *short-segment* limit, cannot be justified on physical grounds. In this respect and to some extent with regards to the influence of dopant[18], the plastic behaviour of sapphire in constant strain rate in basal and prism plane slip differs consistently from the case of silicon[22].

**ACKNOWLEDGEMENTS**

The work was in part supported by the project MAT2003-04199-CO2-02 del Ministerio de Ciencia y Tecnologia (Government of Spain).

**APPENDIX**

**THE STRUCTURE OF DISLOCATIONS AND THE PEIERLS RELIEF IN THE SAPPHIRE LATTICE**

The parameters of the kink pair model ought to be consistent with properties of dislocations in sapphire as inferred from crystallographic considerations. The α-Al$_2$O$_3$ structure is often regarded as a hexagonal close-packed oxygen sub-lattice (…ABABA... stacking, with an average interanionic distance of 0.275 nm) whose octahedral sites are occupied, under the ratio 2/3, by Al ions under a …αβγαβγ… stacking (Figs. A1 and A2). In the real structure, each Al layer (α, β or γ) is actually comprised of two planes separated by 0.052 nm (Ref. 8), that correspond to the displacement of Al ions toward a vacant octahedral site, which in turn introduces a distortion in the oxygen sub-lattice. Considering the real structure is instrumental in discussing basal slip and twinning in α-Al$_2$O$_3$ (Ref. 8). We recall that the lattice parameters are 0.475 nm along the 1/3<$\bar{2}$110> direction and 1.3 nm along the [0001] direction.

In analyzing dislocation properties in post-mortem deformation microstructures by TEM, one currently assumes that the dislocations remaining in the foils at the end of the test are those which exhibited the lowest velocity (or no velocity at all), that is, the rate-controlling dislocations. In sapphire deformed under basal slip at 800°C, dislocations lie primarily along <$\bar{1}$010> directions (Burgers vector at either 30° or 90° from the line).[3] On the other hand in ruby deformed at 1000°C again under basal slip, dislocations are further aligned with <$\bar{2}$110> (Burgers vector at 60° from the line).[18] Mostly screw and edge dislocations have been observed in sapphire deformed under prism slip at 400°C, (Ref. 19) as well as in polycrystals deformed at 630°C (Ref. 23). The following discussion is focused on these dislocation characters.

The strength and directionality of chemical bonds is reflected in the primary and secondary Peierls potentials which, in sapphire, are responsible for the very high stresses that oppose dislocation glide.[3-6] We recall that $h$ and $a$ refer to lattice periodicities experienced by a rectilinear dislocation globally moving in its slip plane in a direction perpendicular to its line, and by a kink moving along a crystallographic direction in the secondary potential. Possible values of $h$ and $a$ are shown in Table AI for various types of dislocations. Unit translations in sapphire (Figs. A1 and A2) are unlikely to account for the Peierls potential periodicity since they may amount to as much as three oxygen-oxygen distances in the basal plane (3 x 0.275 $\cong$ 0.822 nm) or the unrealistic six layers of oxygen (1.3 nm) along the [0001] direction. The period $h$ can be alternatively thought of as corresponding with the various positions of a straight dislocation segment in the potential of atoms (Figs. A1 and A2). In the basal plane, the atomic environment around an edge dislocation (line $a$ in Fig. A1), is not only unchanged after a translation over **b** (position $a_2$) but also after a translation over **b**/2 (line $a_1$), hence $h$ can be reduced to $b/2$ (Table AI). A translation over **b** brings a 60° dislocation (line $b$ in Fig. A1) to $b_1$ at 0.41 nm from the initial position. The two positions located at 1/3 and 2/3 along this path are not substantially different from $b$ and $b_1$ (Fig. A1), suggesting that the potential felt by the line might exhibit a period of $h = 0.41/3 = 0.14$ nm (Table AI). For the 30° dislocation in the basal plane (lines $c$ and $c_1$ in Fig. A1), the only reasonable $h$ corresponds with the lattice translation (Table AI).

In the prism plane, an edge dislocation (line $e$ in Fig. A2) can be translated by **b**/2 with no change in its atomic environment. This holds true after a **b**/3 translation, which corresponds to the dissociated dislocation case,[8] and possibly after the **b**/6 translation as reported in Table AI. A similar situation prevails for a screw dislocation (line $d$ in Fig. A2) gliding along the [0001] direction with $h$ probably not larger than 0.65 nm and possibly as small as $1.3 / 6 = 0.217$ nm (Table AI).

In the absence of a numerical knowledge of the relevant Peierls potential profiles, it is impossible to decide which estimate of $h$ is the most likely while the ratio between the minimum and the maximum $h$ is up to 6 in some cases (Table AI).

Evidently, the above discussion on $h$ holds for $a$, the period of the secondary Peierls potential, because the directions selected above for infinitely straight dislocations should be equally favourable for kinks. A difference may arise though from the fact that kinks have a restricted length, $h$, which as seen above is itself not well-known, not to mention a possible rounding at corners. The situation is thus even more uncertain for $a$ than for $h$, but this does not necessarily constitute a serious limitation in the frame of the kink pair model in that the strain rate $\dot{\varepsilon}$ is far more dependent on $h$ than on $a$ (Eqs. (1) and (4)). Based on the crystal structure, minimum and maximum values for $a$ are suggested in Table AI. Atomistic simulations are expected to shed light on this question but so far, molecular dynamic (MD) simulations have indicated that the equilibrium configuration of a basal edge dislocation exhibits a heavily distorted core[15] with Al ions tending to move toward the centre of the octahedral sites and the anions substantially displaced from their normal positions. The reasonable agreement between the MD simulation estimate of the dislocation line energy of 48 nJ/m (Ref.15) and that based on elasticity[8] should be mentioned.

In conclusion, the scarcity of useful configurations on the core configuration of gliding dislocations in sapphire and of energy profiles makes it problematic to ascribe accurate estimates to $h$ and $a$. We have set $h = a = 0.275$ nm for basal slip, with $h = 0.65$ nm for prism plane slip, in the numerical fits of the kink pair model with the CRSS data. These parameters are, however, rather flexible, thus allowing a fair degree of freedom to refine the fits.

**References**

[1] B. J Pletka., T. E. Mitchell and A. H. Heuer, "Dislocations Structures in Sapphire Deformed by Basal Slip", J. Am. Ceram. Soc. **57** [9] 388-393 (1974).

[2] J. Castaing, A. Muñoz, D. Gómez-García and A. Domínguez-Rodríguez, "Basal Slip in Sapphire (α - $Al_2O_3$)", Mater. Sc. And Eng. *A* **223** 121-125 (1997).

[3] K. L. D. Lagerlof, A.H. Heuer, J. Castaing, J. P. Rivière and T. E. Mitchell, "Slip and Twinning in Sapphire (α - $Al_2O_3$) ", J. Am. Ceram. Soc. **77** [2] 385-397 (1994).

[4] J. Castaing, J. Cadoz and S.H. Kirby, "Deformation of $Al_2O_3$ Single Crystals between 25º and 1800º: Basal and Prismatic Slip", J. Physique *C3* **42** [6] 43-47 (1981).

[5] J. Castaing, J. Cadoz and S. H. Kirby, "Prismatic Slip of $Al_2O_3$ Single Crystals below 1000º in Compression under Hydrostatic Pressure", J. Am. Ceram. Soc. **64** [6] 504-511 (1981).

[6] J. Castaing, A. He, K. P. D. Lagerlof and A. H. Heuer, "Deformation of Sapphire (α - $Al_2O_3$) by basal slip and Basal twinning below 700 ºC", Phil. Mag. **84** [11] 1113-1125 (2004).

[7] J. Cadoz, J. Castaing, D. S. Phillips, A. H. Heuer and T. E. Mitchell, "Work Hardening and Recovery in Sapphire (α-$Al_2O_3$) Undergoing Prism Plane Deformation", Acta Metall. **30** 2205-2218 (1992).

[8] A.H. Heuer, K.P.D. Lagerlof and J.Castaing, "Slip and Twinning Dislocations in Sapphire", Phil. Mag. A **78** 747-763 (1998).

[9] J. P. Hirth and J. Lothe, *Theory of Dislocations*. (Wiley, New York, 1982).

[10] T. E. Mitchell, P. Peralta and J. P. Hirth, Acta Mater., "Deformation by a Kink Mechanism in. High Temperature Materials", **47** 3687 (1999).

[11] T. E. Mitchell, P. M. Anderson, M. I. Baskes, S. P. Chen, R. G. Hoagland and A. Misra, "Nucleation of Kink Pairs on Partial Dislocations: A New Model for Solution Hardening and Softening", Phil. Mag. **83** [11] 1329-1346 (2003).

[12] T. E. Mitchell and A. H. Heuer, "Dislocations and Mechanical Properties of Ceramics" in *Dislocation in Solids*, edited by F. R. N. Nabarro and J. P. Hirth (Elsevier B. V., 2004), chapter 68.

[13] A. Nakamura, T. Yamamoto and Y. Ikuhara, "Direct Observation of Basal Dislocations in Sapphire by HRTEM", Acta Mater. **50** 101-108 (2002).

[14] M. H. Jhon, A. M. Glaeser and D. C. Chzan, "Computational Study of Stacking Faults in Sapphire Using Total Energy Methods", Phys. Rev. *B* **71** 214101 (2005).

[15] C. T. Bodur, J. Chang and A. S. Argon, "Molecular Dynamics Simulations of Basal and Pyramidal System Edge Dislocations in Sapphire", J. Europ. Ceram. Soc. **25** 1431-1439 (2005).

[16] B. J. Pletka, A. H. Heuer and T. E. Mitchell, "Work Hardening in Sapphire ($\alpha$ - $Al_2O_3$)", Acta Metal. **25** 25-33 (1977).

[17] A. Moulin, M. Condat and L. P. Kubin, "Mesoscale Modelling of the Yield Point Properties of Silicon Crystals". Acta Mater. **47** [10] 2879-2888 (1999).

[18] M. Castillo Rodríguez, A. Muñoz, J. Castaing, P. Veyssière and A. Domínguez Rodríguez, "Chromium Hardening and Peierls Mechanism for Basal Slip in Sapphire ($\alpha$ - $Al_2O_3$) at Temperatures between 900 and 1500 ºC", J. Europ. Ceram. Soc. **27** 3317-3324 (2007).

[19] J. Cadoz, J. P. Rivière and J. Castaing, in *Deformation of ceramic materials II*, edited by R.E. Tressler and R.C Bradt (Plenum Press, New York, 1984), p. 213.

[20]J. Cadoz, D. Hokim, M. Meyer and J. P. Rivière, "Observation of Dislocations Associated with Prism Plane Slip in Alumina Single Crystals", Revue Phys. Appl. **12** 473-481 (1977).

[21]J. Castaing, P. Veyssière, L. Kubin and J. Rabier, "The Plastic Deformation of Silicon between 300°C and 600°C", Phil. Mag. *A* **44** 1407-1413 (1981).

[22]L. P. Kubin and B. Devincre, *Mesoscale simulations of dislocations and plasticity* (in preparation).

[23]S. Lartigues Korinek and J. Castaing, "Slip and Twinning in Polycrystalline Alumina (α-$Al_2O_3$) Deformed under Hydrostatic Pressure between 600° and 1000°C", J. Amer. Ceram. Soc. **86** 566-573 (2003).

[24]J. Cadoz, J. Castaing and J. Philibert, "Glissement Prismatique de $Al_2O_3$ Déformé par Compression", Revue Phys. Appl. **16** 135-144 (1981).

[25]T. E. Mitchell, J. P. Hirth and A. Misra, "Apparent Activation Energy and Stress Exponent in Materials with a High Peierls Stress", Acta Mater. **50** 1087-1093 (2002).

**Figure captions**

Figure 1. Plot of CRSS (logarithmic scale) versus temperature for basal and prism plane slips. The origins of the different experimental data points are indicated. Note that straight lines are drawn through the experimental points.

Figure 2. Graph of CRSS versus temperature for basal slip and for prism plane slip showing both experimental data and the curves from the fits using the *long-segment* limit (Eq. (1)).

Figure 3. Graph of CRSS versus temperature for basal slip and for prism plane slip showing both experimental data and the curves from the fits using the *short-segment* limit (Eq. (4)).

Figure A1. Basal plane for sapphire looking along [$000\bar{1}$]. The large circles are for oxygen stacked as … ABABAB …; the anions are shown for A (dark) and B (clear) layers. The cations have three different sizes according to the α, β or γ layer where they are located. The lines *a*, *b* and *c* correspond to dislocations in the basal plane.

Figure A2. Prism plane of sapphire viewed along [$\bar{2}110$]. The …ABAB… stacking for oxygen and …αβγαβγ… stacking for Al are shown. The lines *d* and *e* correspond to dislocation segments for slip in the prism plane.

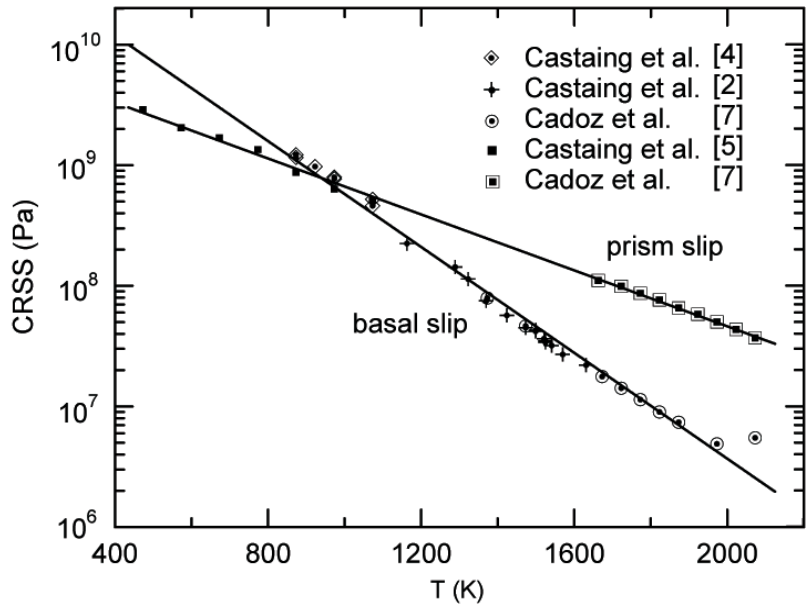

Figure 1. Plot of CRSS (logarithmic scale) versus temperature for basal and prism plane slips. The origins of the different experimental data points are indicated. Note that straight lines are drawn through the experimental points.

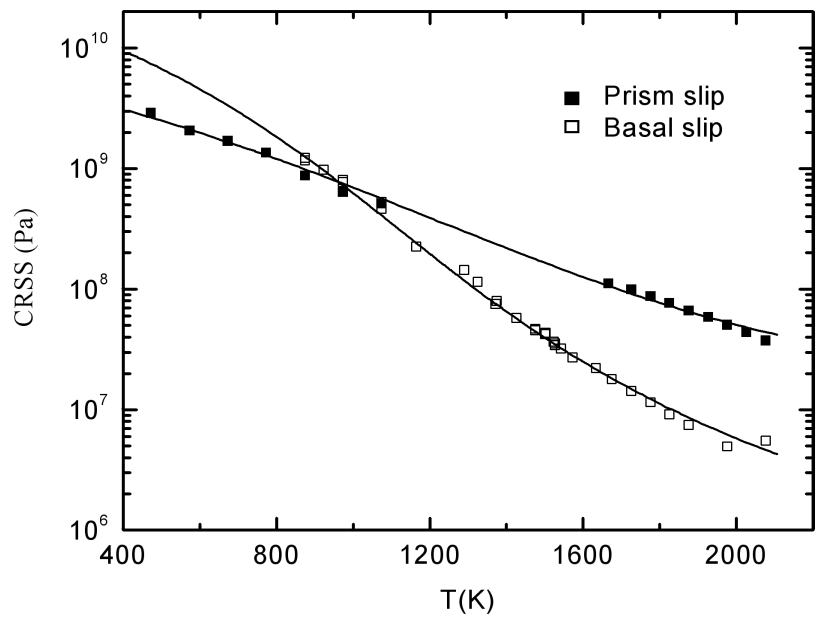

Figure 2. Graph of CRSS versus temperature for basal slip and for prism plane slip showing both experimental data and the curves from the fits using the *long-segment* limit (Eq. (1)).

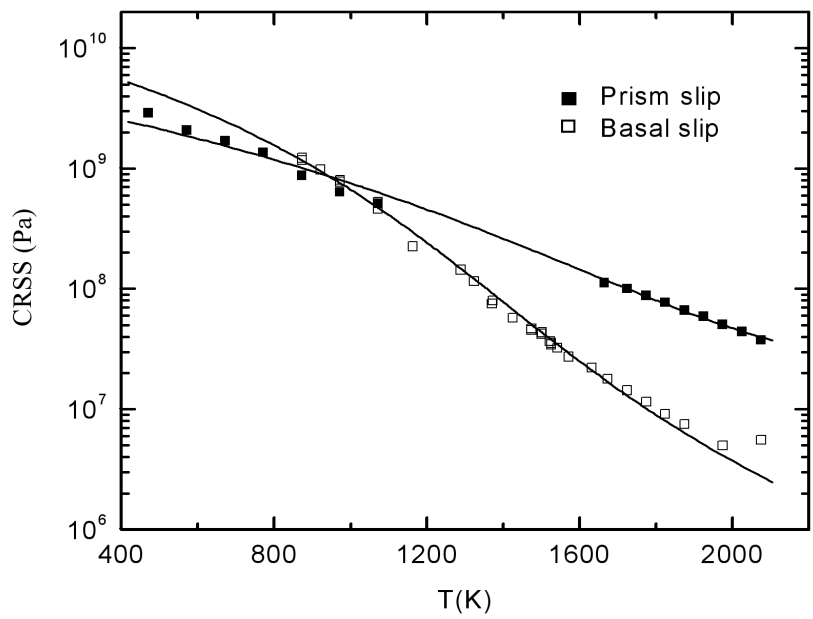

Figure 3. Graph of CRSS versus temperature for basal slip and for prism plane slip showing both experimental data and the curves from the fits using the *short-segment* limit (Eq. (4)).

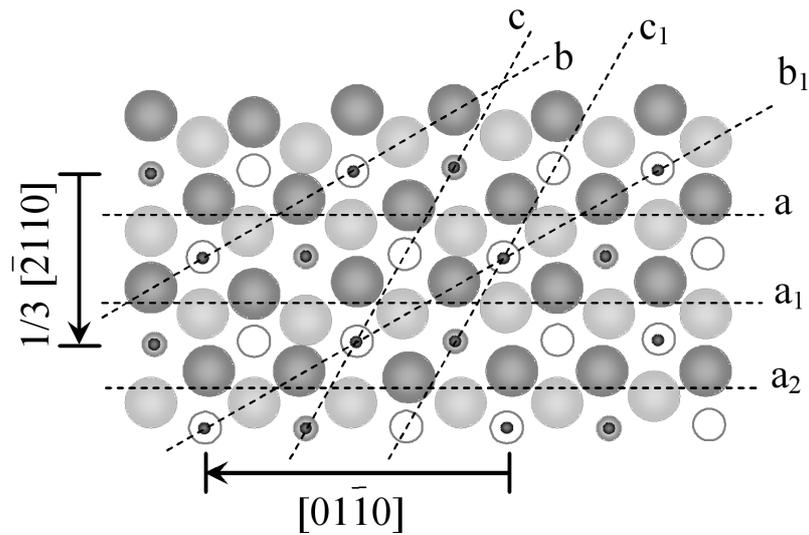

Figure A1. Basal plane for sapphire looking along $[000\bar{1}]$. The large circles are for oxygen stacked as … ABABAB …; the anions are shown for A (dark) and B (clear) layers. The cations have three different sizes according to the α, β or γ layer where they are located. The lines *a*, *b* and *c* correspond to dislocations in the basal plane.

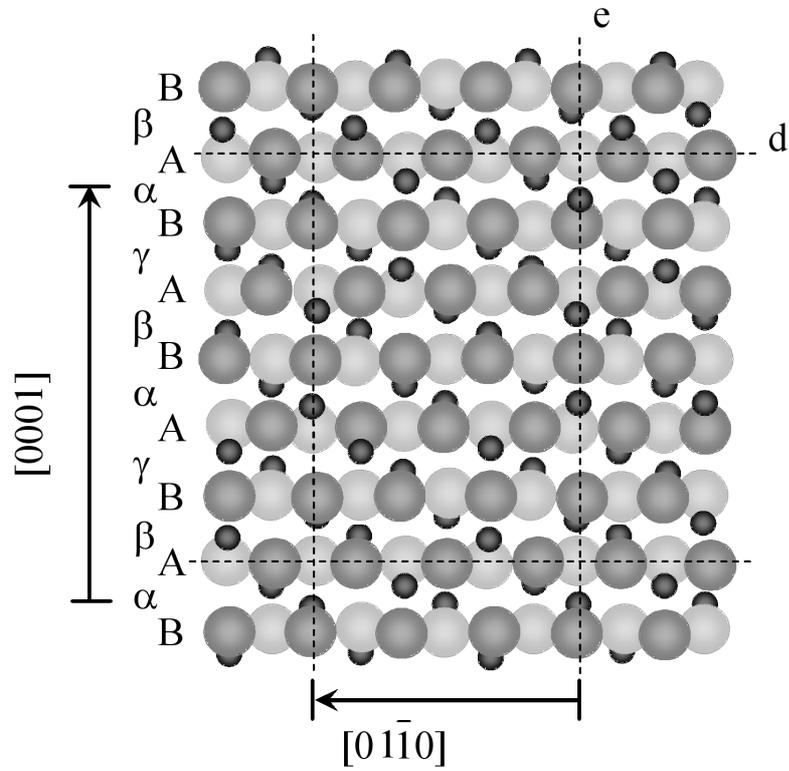

Figure A2. Prism plane of sapphire viewed along [$\bar{2}110$]. The …ABAB… stacking for oxygen and …αβγαβγ… stacking for Al are shown. The lines *d* and *e* correspond to dislocation segments for slip in the prism plane.

Table I. Values obtained for the various parameters of the kink model for different fits of CRSS data of basal slip in sapphire. The symbol * indicates that the quantity was ascribed a fixed value to perform the fit.

| Parameters | | Long-segment limit | | Short-segment limit | |
|---|---|---|---|---|---|
| | | I | II | III | IV |
| $\mu$ (GPa) | | 156* | 115 | 156* | 115* |
| $L$ (μm) | | ---- | ---- | 8.9 | 1.2 |
| $C$ | | 0.054 | 0.108 | 0.108* | 0.121* |
| $\rho$ (m$^{-2}$) | $T = 1873$ K, $\sigma = 7.4$ MPa | $3.4 \cdot 10^{12}$ | $8.6 \cdot 10^{11}$ | $8.6 \cdot 10^{11}$ | $6.8 \cdot 10^{11}$ |
| | $T = 973$ K, $\sigma = 783$ MPa | $3.8 \cdot 10^{16}$ | $9.6 \cdot 10^{15}$ | $9.6 \cdot 10^{15}$ | $7.6 \cdot 10^{15}$ |
| $Q_D + F_K$ (eV) | | 3.6 | 3.4 | ---- | ---- |
| $Q_D + 2 F_K$ (eV) | | ---- | ---- | 5.0 | 4.6 |

Table II. Parameters resulting from different fits for prism plane slip in sapphire.

| Parameters | | Long-segment Limit (I) | Short-segment Limit (II) |
|---|---|---|---|
| $\mu$ (GPa) | | 194 | 177 |
| $L$ (μm) | | ---- | 1.8 |
| C | | 2.8 | 3.04 |
| $\rho$ (m$^{-2}$) | $T = 1873$ K, $\sigma = 65.5$ MPa | $1.9 \cdot 10^{11}$ | $2 \cdot 10^{11}$ |
| | $T = 973$ K, $\sigma = 635.0$ MPa | $1.8 \cdot 10^{13}$ | $1.9 \cdot 10^{13}$ |
| $Q_D + F_K$ (eV) | | 3.5 | ---- |
| $Q_D + 2 F_K$ (eV) | | ---- | 5.4 |

Table AI. Estimates of the periods *h* and *a* of the primary and secondary Peierls potentials involved in the kink pair motion in the basal and prism planes. For each type of dislocation, the values of the repeat distances in the lattice and the shortest values deduced from examination of the sapphire lattice are given.

| Dislocation | *h* (nm) | Comment | *a* (nm) | Comment |
|---|---|---|---|---|
| Edge; $\langle 01\bar{1}0 \rangle$ line<br>b // $\langle \bar{2}110 \rangle$ | 0.475<br>0.237 | Lattice translation ( b )<br>b / 2 | 0.822<br>0.137 | Lattice translation<br>0.822 / 6 |
| 60°; $\langle 1\bar{2}10 \rangle$ line<br>b // $\langle \bar{2}110 \rangle$ | 0.41<br>0.14 | Lattice translation<br>0.41 / 3 | 0.475<br>0.158 | Lattice translation<br>0.475 / 3 |
| 30°; $\langle \bar{1}100 \rangle$ line<br>b // $\langle \bar{2}110 \rangle$ | 0.237 | Lattice translation | 0.822<br>0.137 | Lattice translation<br>0.822 / 6 |
| Edge; $[0001]$ line<br>b // $\langle 01\bar{1}0 \rangle$ | 0.822<br>0.411<br>0.137 | Lattice translation ( b )<br>b / 2<br>b / 6 | 1.3<br>0.217 | Lattice translation<br>1.3 / 6 |
| Screw ; line and<br>b // $\langle 01\bar{1}0 \rangle$ | 1.3<br>0.65<br>0.217 | Lattice translation<br>1.3 / 2<br>Oxygen plane distance | 0.822<br>0.137 | Lattice translation<br>0.822 / 6 |